\documentclass[
    ,final            
  ]
  {aipproc}

 \layoutstyle{6x9}
\usepackage{color}

\newcommand {\blue} {\color{blue}}

\usepackage{epsfig}
 \usepackage{rotating}
 \usepackage{amsmath,amssymb}
 \usepackage{graphicx}
\usepackage{ifthen,array}
 \usepackage{setspace}
 \usepackage{fancyhdr}
 \usepackage{moreverb}
 \usepackage{rotating}
\usepackage{amsfonts}
\usepackage{bbm}

\newcommand{\Sol}  {\textsc{sol}}

\newcommand{\Atm}  {\textsc{atm}}

\newcommand{\Dmq}  {\Delta m^2}
\newcommand{\Dms}  {\Delta m^2_{21}}
\newcommand{\Dma}  {\Delta m^2_{31}}

\def\e6{$E(6)$}
\def\10{$SO(10)$}
\def\21{$SU(2) \otimes U(1) $}

\def\422{$SU(4) \otimes SU(2) \otimes SU(2)$}
\def\321{$SU(3) \otimes SU(2) \otimes U(1)$}
\def\lsim{\raise0.3ex\hbox{$\;<$\kern-0.75em\raise-1.1ex\hbox{$\sim\;$}}}
\def\gsim{\raise0.3ex\hbox{$\;>$\kern-0.75em\raise-1.1ex\hbox{$\sim\;$}}}
\def\lfv{lepton flavour violation }

\def\meff{\langle m_{\nu} \rangle}
\newcommand{\ed}{\end{document}}
\DeclareMathAlphabet{\mathsc}{OT1}{cmr}{m}{sc}

\def \znbb {$0\nu\beta\beta$ }

\def\meff{\langle m_{\nu} \rangle}

\let\vev\VEV

\def\e6{$E(6)$}
\def\10{$SO(10)$}
\def\21{$SU(2) \otimes U(1) $}

\def\422{$SU(4) \otimes SU(2) \otimes SU(2)$ }
\def\321{$SU(3) \otimes SU(2) \otimes U(1)$ }

\renewcommand{\baselinestretch}{1.01}

\newcommand{\AddrAHEP}{%
 Instituto de F\'{\i}sica Corpuscular,
  C.S.I.C. -- Universitat de Val{\`e}ncia \\
  Edificio de Institutos de Paterna, Apartado 22085,
  E--46071 Val{\`e}ncia, Spain\\}

\begin{document}
\newcommand{\Od}{{\cal O}}

\title{Neutrinos as cosmic messengers}

\author{J. W. F. Valle}{
address={\AddrAHEP}}

\begin{abstract}

  I briefly review the current status of neutrino oscillation
  parameters and discuss the role of neutrinos as cosmological probes,
  that could possibly induce the baryon asymmetry as well as the dark
  matter in the Universe. I comment on the origin of neutrino masses
  in seesaw-type and low-scale models and mention some of their
  laboratory signals.

\end{abstract}
\maketitle


\bibliographystyle{aipproc}   

 \section{Status of neutrino oscillations}
\label{sec:stat-neutr-oscill}

The discovery of neutrino oscillations marks a turning point in
particle physics as it implies that neutrinos have masses. Given their
weak interaction, neutrinos play a special role as cosmic probes as
they may provide information about very early stages of the evolution
of the Universe. Last, but not least, understanding their properties
may provide a valuable clue of what may lie ahead of the Standard
Model (SM) of basic interactions.

Here I summarize the status of the neutrino oscillation results after
the Neutrino 2008 conference~\cite{Neutrino2008}, given in
Ref.~\cite{Schwetz:2008er}. Evidence for neutrino oscillations coming
from ``celestial'' (solar and atmospheric) neutrinos is unambiguously
confirmed by ``laboratory'' neutrinos produced at reactors and
accelerators.
The basic theoretical layout for the description of neutrino
oscillation data has been given almost thirty years ago and involves
the concept of the lepton mixing matrix, the lepton analogue of the
quark mixing matrix. In its simplest unitary 3-dimensional form is
given as~\cite{schechter:1980gr}
\begin{equation}
  \label{eq:2227}
K =  \omega_{23} \omega_{13} \omega_{12}
\end{equation}
where each $\omega$ is effectively $2\times 2$, characterized by an
angle and a basic Majorana CP phase present already with two
generations of neutrinos~\cite{schechter:1980gr}.  These do not affect
oscillations~\cite{maj-phase:1981gk}, moreover, \underline{current}
neutrino oscillation data have no sensitivity to the remaining Dirac
CP violation phase.  Thus we set all three phases to zero.
In this approximation oscillations depend on the three mixing
parameters $\sin^2\theta_{12}, \sin^2\theta_{23}, \sin^2\theta_{13}$
and on the two mass-squared splittings $\Dms \equiv \Delta m^2_{21}
\equiv m^2_2 - m^2_1$ and $\Dma \equiv \Delta m^2_{31} \equiv m^2_3 -
m^2_1$ characterizing solar and atmospheric transitions.  The
hierarchy $\Dms \ll \Dma$ implies that, to a good approximation, one
can set $\Dms = 0$ in the analysis of atmospheric and accelerator
data, and $\Dma$ to infinity in the analysis of solar and reactor
data.

The analysis of the data requires accurate calculations of solar and
atmospheric fluxes, neutrino cross sections and response functions, as
well as a careful description of neutrino propagation in the Sun and
the Earth, taking into account matter effects.

\begin{figure}[t]
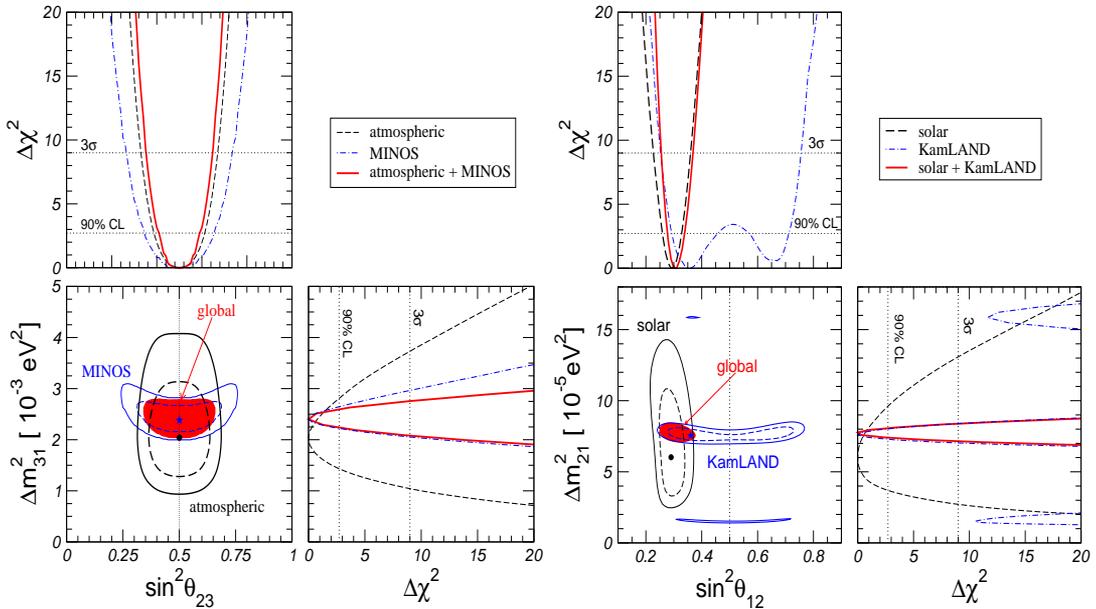
 \centering
\includegraphics[width=.48\linewidth,height=8cm]{F-atm-min-08-3pan.eps}
\includegraphics[width=.48\linewidth,height=8cm]{F-sol-kl-08-3pan.eps}
\caption{\label{fig:global} %
  Neutrino oscillation parameter regions after Neutrino 2008
  conference, from Ref.~\cite{Schwetz:2008er}. }
\end{figure}

\begin{figure}[t]
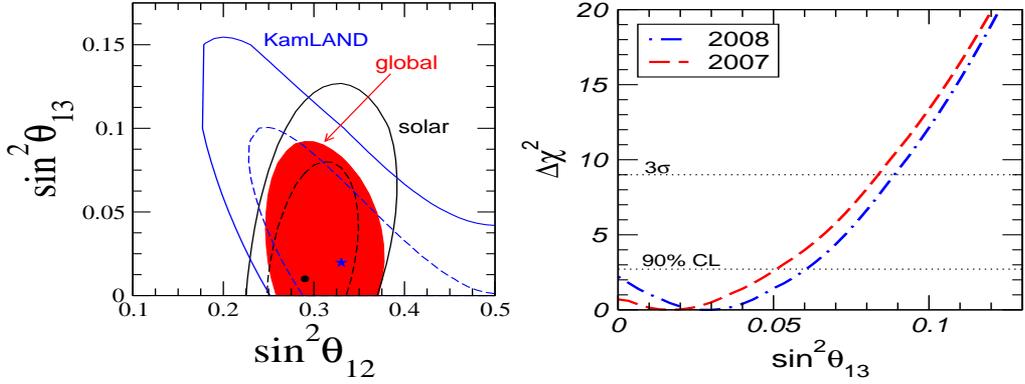
 \centering
\includegraphics[height=5cm,width=.45\linewidth]{s12-s13-tension.eps}
\includegraphics[height=5cm,width=.45\linewidth]{th13-sol-07vs08-lin.eps}
\caption{\label{fig:th13}%
Constraints on $\sin^2\theta_{13}$ from different parts of
  the global data given in Ref.~\cite{Schwetz:2008er}.}
\end{figure}

The resulting three--neutrino oscillation parameters obtained in the
global analysis are summarized in Figs.~\ref{fig:global} and
\ref{fig:th13}. The analysis includes all new neutrino oscillation
data, as of the recent Neutrino 2008 conference~\cite{Schwetz:2008er}.
These include the data released this summer by the MINOS
collaboration, the data of the neutral current counter phase of the
SNO solar neutrino experiment, as well as the latest KamLAND and
Borexino data.

The left panel gives the leading ``atmospheric'' oscillation
parameters $\theta_{23}$ \& $\Dma$ from the interplay of data from
artificial and natural neutrino sources. We show $\chi^2$-profiles and
allowed regions at 90\% and 99.73\%~CL (2~dof) for atmospheric and
MINOS, as well as the 99.73\%~CL region for the combined analysis
(including also K2K). The dot, star and diamond indicate the best fit
points of atmospheric data, MINOS and global data, respectively. We
minimize with respect to $\Dms$, $\theta_{12}$ and $\theta_{13}$,
including always solar, KamLAND, and CHOOZ data.
 
The right panel gives the corresponding ``solar'' oscillation
parameters $\theta_{12}$ \& $\Dms$ obtained by combining solar and
reactor neutrino data. We show $\chi^2$-profiles and allowed regions
at 90\% and 99.73\%~CL (2~dof) for solar and KamLAND, as well as the
99.73\%~CL region for the combined analysis. The dot, star and diamond
indicate the best fit points of solar data, KamLAND and global data,
respectively. We minimize with respect to $\Dma$, $\theta_{23}$ and
$\theta_{13}$, including always atmospheric, MINOS, K2K and CHOOZ
data.

The angle $\theta_{13}$ holds the key for future searches for CP
violation in neutrino oscillations.
Fig.~\ref{fig:th13} summarizes the information on $\theta_{13}$ from
present data, the right panel compares the situation in 2007 and
2008. One sees that the current data slightly prefer a nonzero value
for $\theta_{13}$, though this is not significant and we interpret
this as giving a bound on $\theta_{13}$. An important contribution to
this bound comes, of course, from the CHOOZ reactor experiment
combined with the determination of $|\Dmq_{31}|$ from atmospheric and
long-baseline experiments. The complementarity of different data sets
provides a non-trivial constraint on $\theta_{13}$,
namely~\footnote{Note: the bounds in Eq.~(\ref{eq:th13}) are given
  for 1~dof, while the regions in Fig.~\ref{fig:th13} (left) are
  90\%~CL for 2~dof}:
\begin{equation}\label{eq:th13}
  \sin^2\theta_{13} \le \left\lbrace \begin{array}{l@{\qquad}l}
      0.060~(0.089) & \text{(solar+KamLAND)} \\
      0.027~(0.058) & \text{(CHOOZ+atm+K2K+MINOS)} \\
      0.035~(0.056) & \text{(global data)}
    \end{array} \right.
\end{equation}

Within a three--neutrino scheme CP violation disappears when two
neutrinos become degenerate or when one of the angles
vanishes~\cite{schechter:1980bn}.  As a result CP violation is doubly
suppressed, first by the small ratio $\alpha \equiv \Dms/\Dma$ of the
two mass-squared differences, and also by the small value of
$\theta_{13}$.
There is now an ambitious long-term effort towards probing CP
violation in neutrino oscillations in long-baseline
experiments~\cite{Bandyopadhyay:2007kx,Nunokawa:2007qh}.  The current
status of the determination of the parameter $\alpha$ is
\begin{equation}
  \alpha \equiv \frac{\Delta m^2_{21}}{|\Delta m^2_{31}|} = 
0.032\,, \quad 0.027 \le \alpha \le 0.038 \quad (3\sigma) \,,
\end{equation}
%

%


The growing precision of oscillation experiments also opens good
prospects for probing small effects beyond Eq.~(\ref{eq:2227}) such as
unitarity violation and other forms of non-standard neutrino
interactions~\cite{schechter:1980gr}.
Here I wish to stress that reactor neutrino data play a crucial role
in testing the robustness of solar oscillations vis a vis
astrophysical uncertainties, such as magnetic fields in the solar
radiative~\cite{Burgess:2003su,Burgess:2003fj,burgess:2002we} or
convective zone~\cite{miranda:2000bi,guzzo:2001mi,barranco:2002te},
leading to stringent limits on neutrino magnetic transition
moments~\cite{Miranda:2004nz}.
KamLAND has also played a key role in identifying oscillations as
``the'' solution to the solar neutrino problem~\cite{pakvasa:2003zv}
and also in pinning down the large-mixing-angle oscillation solution
among the previous wide range of
possibilities~\cite{gonzalez-garcia:2000sq}.

\begin{figure}[t] \centering
    \includegraphics[height=3cm,width=.45\linewidth]{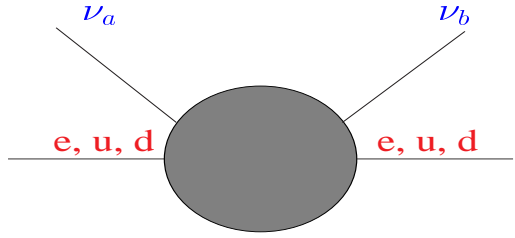}
    \caption{\label{fig:nuNSI} %
      Non-standard neutrino interactions arise, e.~g., from the
      non-unitary structure of charged current weak interactions
      characterizing seesaw-type schemes~\cite{schechter:1980gr}.}
\end{figure}

However, there is still an ambiguity left in the interpretation of the
solar data in the presence of non-standard neutrino interactions
(NSI), illustrated in Fig.~\ref{fig:nuNSI}.  Indeed, most neutrino
mass generation mechanisms imply the existence of such dimension-6
operators, typically sub-weak strength ($\sim \varepsilon G_F$). They
can be of two types: flavour-changing (FC) and non-universal
(NU). Their presence leads to the possibility of resonant neutrino
conversions even in the absence of neutrino
masses~\cite{valle:1987gv}.
For example, in the inverse seesaw
model~\cite{mohapatra:1986bd,Deppisch:2004fa} the non-unitary piece of
the lepton mixing matrix can be sizeable, hence the induced
non-standard interactions.  Relatively sizable NSI strengths may also
be induced in supersymmetric unified models~\cite{hall:1986dx} and
models with radiatively induced neutrino
masses~\cite{zee:1980ai,babu:1988ki}.

Although first determinations of atmospheric neutrino data allowed for
an NSI interpretation~\cite{GonzalezGarcia:1998hj}, thanks to the
large currently available statistics of data over a wide energy range
the determination of $\Dma$ and $\sin^2\theta_\Atm$ is now hardly
affected by the presence of NSI, at least within the 2--neutrino
approximation~\cite{fornengo:2001pm}. Future neutrino factories will
substantially improve this bound~\cite{huber:2001zw}.

In contrast, the determination of solar neutrino parameters is not yet
robust against the existence of NSI~\cite{Miranda:2004nb}, even if
reactor data are included. One can show that even a small residual
non-standard interaction may have dramatic consequences for the
sensitivity to $\theta_{13}$ at a neutrino
factory~\cite{huber:2001de}.  Improving the sensitivities on NSI
constitutes a necessary step and opens a window of opportunity for
neutrino physics in the precision age.

\section{Lepton number violation}
\label{sec:lepton-number-lepton}

\begin{figure}[h]
  \centering
\includegraphics[width=6cm,height=4cm]{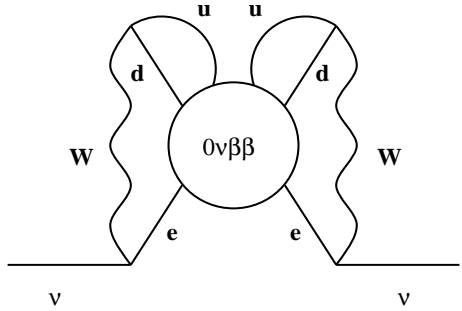}
\caption{Neutrinoless double beta decay and Majorana mass are
  equivalent~\cite{Schechter:1982bd}.}
 \label{fig:bbox}
\end{figure}

Neutrino oscillations are blind to whether neutrinos are Dirac or
Majorana fermions. In contrast, lepton number violating (LNV)
processes, such as \znbb~\cite{Schechter:1982bd}~\footnote{Neutrino
  transition magnetic
  moments~\cite{schechter:1981hw,Wolfenstein:1981rk,pal:1982rm,kayser:1982br}
  provide another example of LNV processes.}, do have the potential of
probing the intrinsic nature of neutrinos.
For example, it will in general be sensitive to CP violation induced
by the so-called Majorana
phases~\cite{schechter:1980gr,maj-phase:1981gk}, inaccessible in
conventional oscillations.
Hence the search for neutrinoless double beta
decay~\cite{Avignone:2007fu} constitutes a major goal for the future.
It has also been argued that, in a gauge theory, \emph{irrespective of
  the mechanism that induces \znbb}, it necessarily implies a Majorana
neutrino mass~\cite{Schechter:1982bd}, as illustrated in Fig.
\ref{fig:bbox}.  Indeed, in this resides the basic theoretical
significance of \znbb.
This is known as the ``black-box'' theorem~\cite{Schechter:1982bd}.
Although the theorem itself holds in any ``natural'' gauge theory, its
quantitative implications are very model-dependent, for a recent
discussion see \cite{Hirsch:2006yk}.

The observation of neutrino oscillations implies that \znbb must be
induced by the exchange of light Majorana neutrinos, through the
so-called \emph{mass-mechanism}. The corresponding amplitude is
sensitive both to the Majorana CP phases~\cite{schechter:1980gr}, and
also to the absolute scale of neutrino mass, neither of which can be
probed in oscillations. Note that the absolute mass scale parameter
probed by neutrinoless double beta decay is complementary to those
probed in high sensitivity beta decay studies~\cite{Drexlin:2005zt},
and observations of the cosmic microwave background and large scale
structure~\cite{Lesgourgues:2006nd}.

Taking into account current neutrino oscillation
parameters~\cite{Schwetz:2008er} and state-of-the-art nuclear matrix
elements~\cite{Rodin:2007fz} one can determine the average mass
parameter $\meff$ characterizing the neutrino exchange contribution to
\znbb, as shown in Fig. 10 of Ref.~\cite{Valle:2006vb}.
Models with quasi-degenerate
neutrinos~\cite{babu:2002dz}~\cite{caldwell:1993kn}~\cite{ioannisian:1994nx}
give the largest \znbb signal. In normal hierarchy models there is in
general no lower bound on $\meff$, since there can be a destructive
interference amongst the neutrino amplitudes (for an exception, see
Ref.~\cite{Hirsch:2005mc}; in that specific model a lower bound on
$\meff$ exists, which depends, as expected, on the value of the
Majorana CP violating phase $\phi_1$).  In contrast, the inverted
neutrino mass hierarchy implies a ``lower'' bound for the \znbb
amplitude.

The best current limit on $\meff$ comes from the Heidelberg-Moscow
experiment. There is also a claim made in
Ref.~\cite{Klapdor-Kleingrothaus:2004wj} (see
also~\cite{Aalseth:2002dt}) which will be important to confirm or
refute in future experiments. GERDA will provide an independent check
of this claim~\cite{Aalseth:2002rf}. SuperNEMO, CUORE, EXO, MAJORANA
and possibly other experiments will further extend the sensitivity of
current \znbb searches~\cite{dbd06}.

\section{Lepton flavor violation}
\label{sec:lfv}

Given that neutrinos and charged leptons sit in the same electroweak
doublet, and \lfv has been shown to occur in neutrino propagation, it
is natural to expect that it may also show up as transitions directly
involving the charged leptons themselves.  Indeed, this occurs in
seesaw-type schemes of neutrino mass, either through neutral heavy
lepton
exchange~\cite{Bernabeu:1987gr,gonzalez-garcia:1992be,Ilakovac:1994kj}
or via supersymmetric
contributions~\cite{Hall:1985dx,borzumati:1986qx,casas:2001sr,Antusch:2006vw}.
Moreover, supersymmetry brings in the possibility of direct \lfv in
the production of supersymmetric particles at the
LHC~\cite{Hirsch:2008dy}.

\begin{figure}[t]
  \centering
\includegraphics[width=6cm,height=5cm]{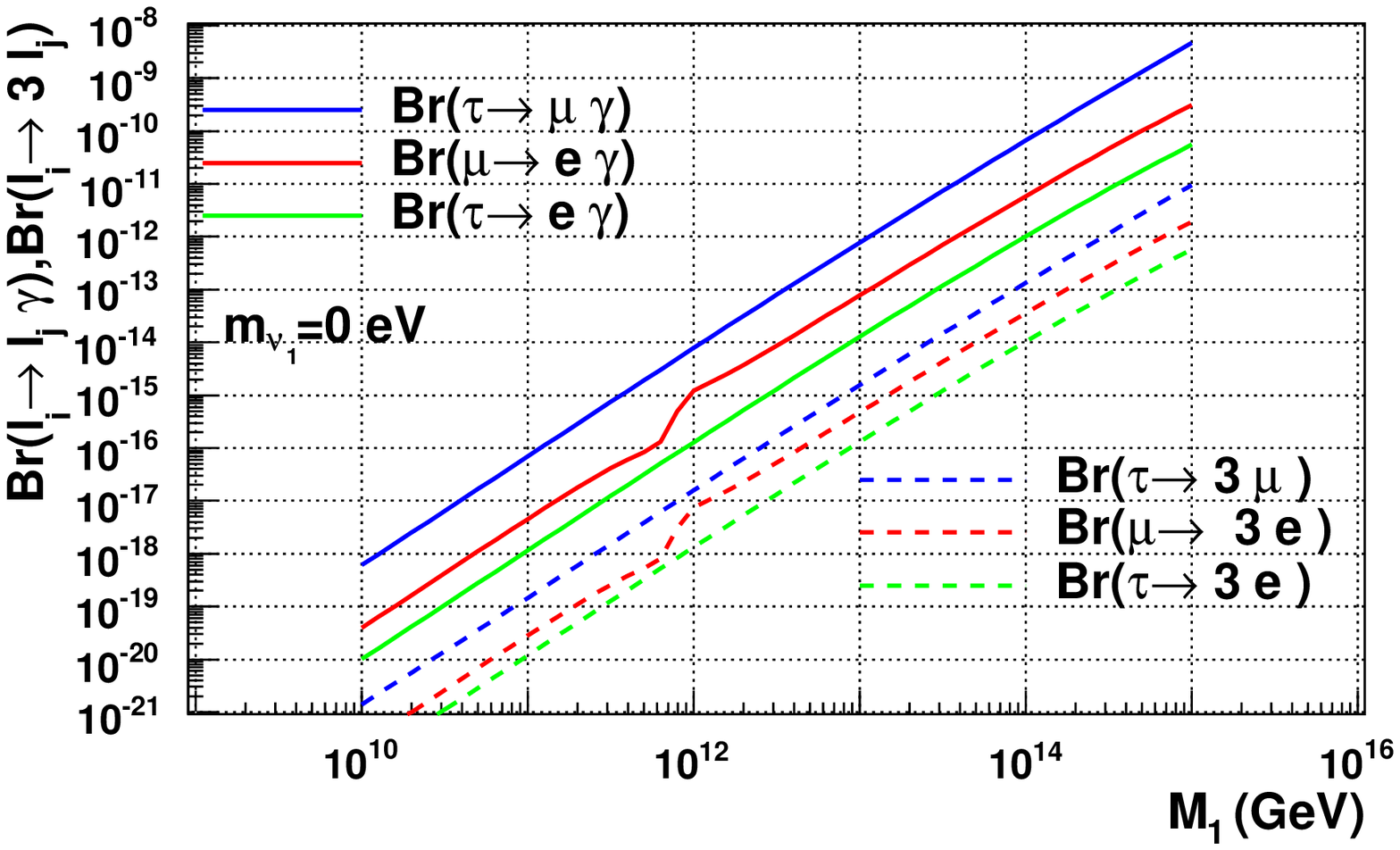}
\includegraphics[width=6cm,height=5cm]{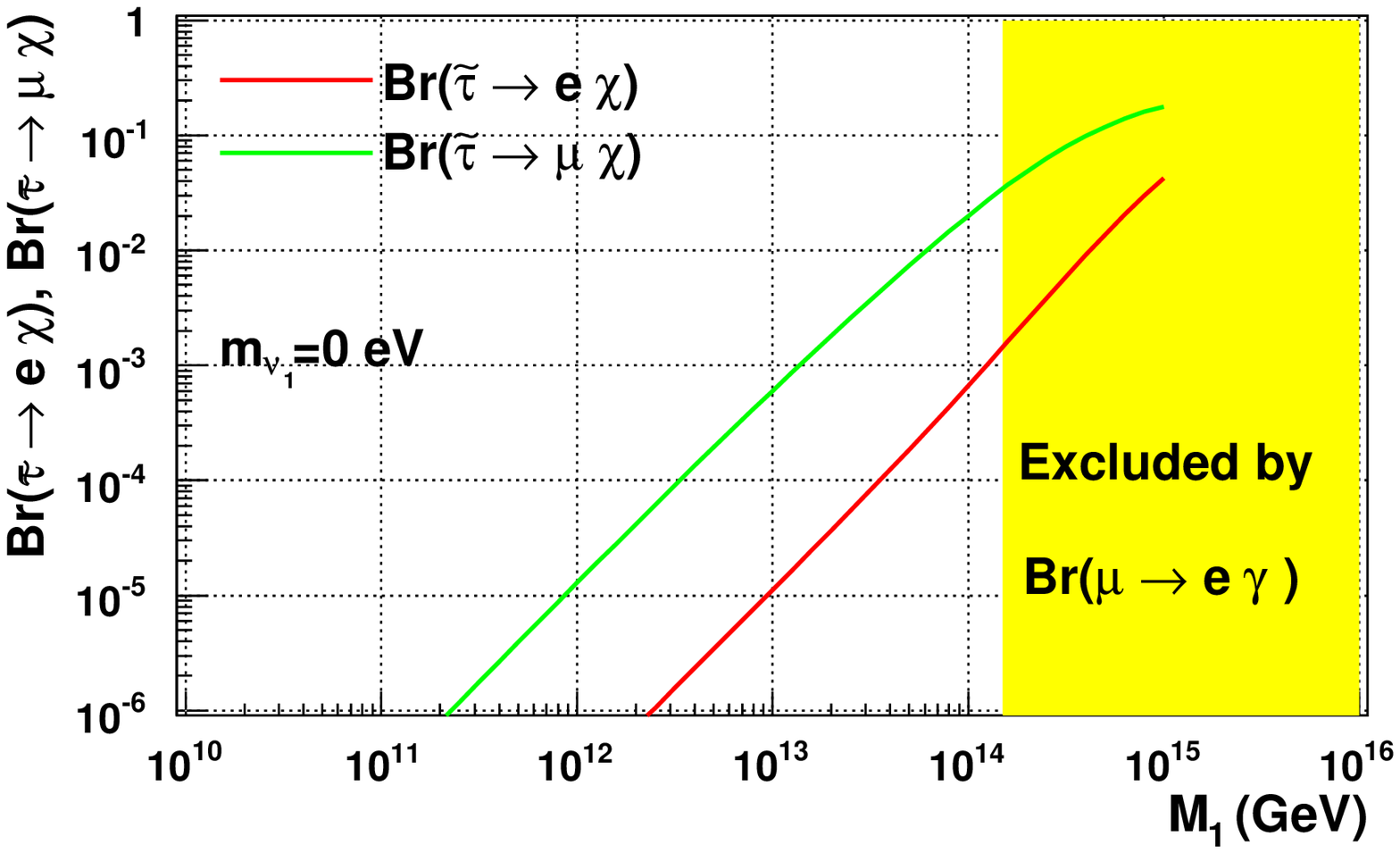}          
\caption{The left panel gives minimal type-I seesaw expectations for
  loop-induced LFV rates, versus the right-handed neutrino mass, while
  the right panel illustrates the possible direct LFV in the decays of
  staus at the LHC, details in Ref.~\cite{Hirsch:2008dy}.}
\label{fig:lfv-seesaw}
\end{figure}

As illustrated in Fig.~\ref{fig:lfv-seesaw} supersymmetry can lead to
sizeable rates for \lfv processes even within the simplest minimal
supergravity version of the type-I seesaw
mechanism~\cite{Hirsch:2008dy}.
The figures illustrate the theoretical branching ratios for the LFV
scalar tau decays, ${\tilde \tau}_2 \to (e,\mu) + \chi^0_1$, as well
as loop-induced LFV decays at low energy, such as $l_i \to l_j +
\gamma$ and $l_i \to 3 l_j$, for given choice of the unknown seesaw
parameters, see Ref.~\cite{Hirsch:2008dy} for details.
One can show that in some simple scenarios for the unknown
right-handed parameters, the ratios of LFV branching ratios correlate
with neutrino oscillation parameters. If the overall mass scale of the
left neutrinos and the value of the reactor angle were known, the
study of LFV allows, in principle, to extract information about the so
far unknown right-handed neutrino parameters.

It is remarkable that, in general, the rates for \lfv processes may be
sizeable, despite the small values of the light neutrino masses
determined in current neutrino experiments. Indeed, an important
theoretical point is that \lfv and CP violation can occur in the
massless neutrino
limit~\cite{Bernabeu:1987gr,branco:1989bn,rius:1990gk,gonzalez-garcia:1992be}.
As a result the allowed rates are \underline{not} suppressed by the
smallness of neutrino masses. In the extended seesaw scheme one can
understand the interplay of both types of contributions. It is
shown~\cite{Deppisch:2004fa} that \(Br(\mu\to e\gamma)\) and the
nuclear $\mu^--e^-$ conversion rates lie within planned sensitivities
of future experiments such as PRISM~\cite{Kuno:2000kd}.
\begin{figure}[!h]
  \centering
\includegraphics[width=7cm,height=5cm]{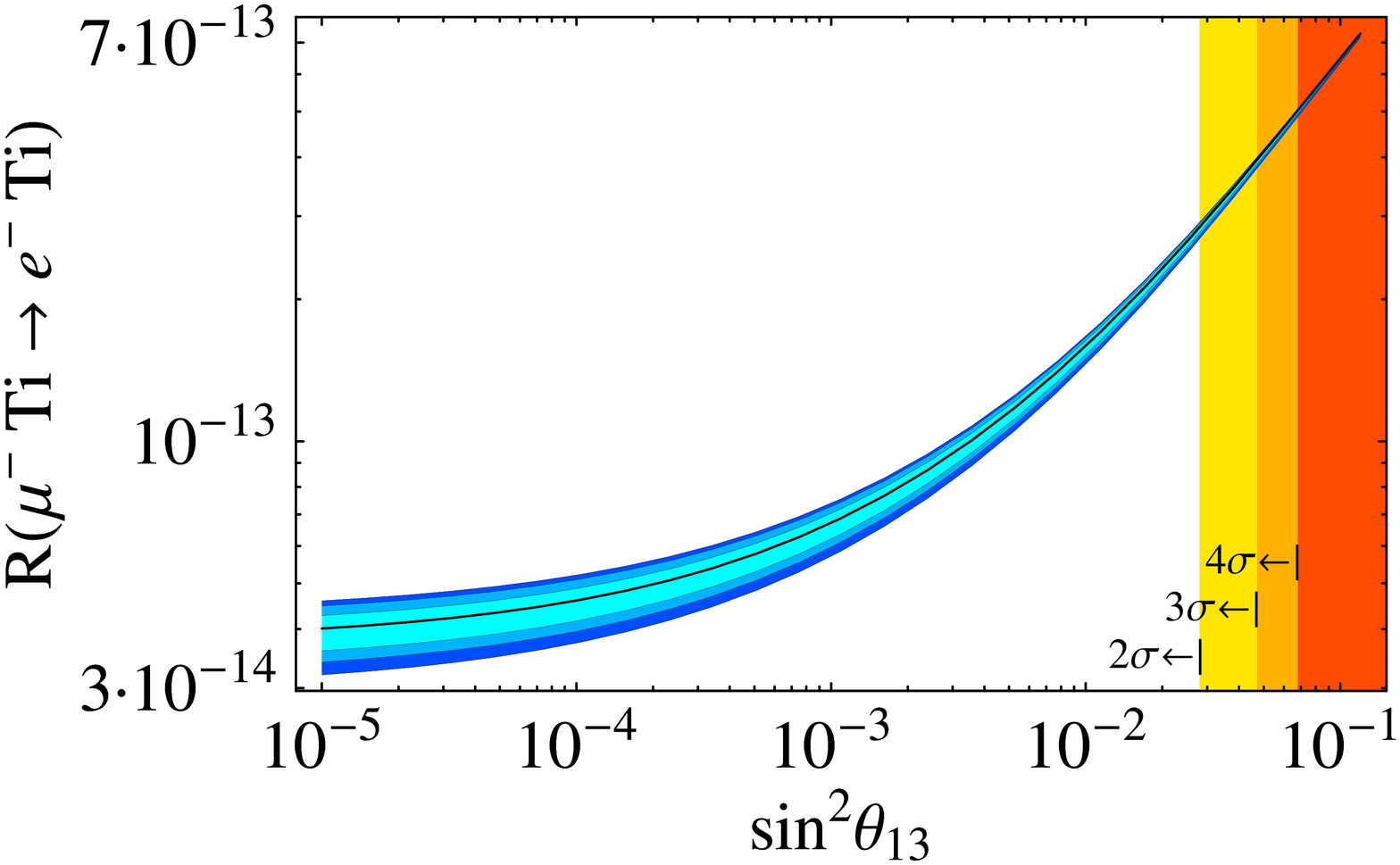}
\includegraphics[width=7cm,height=5cm]{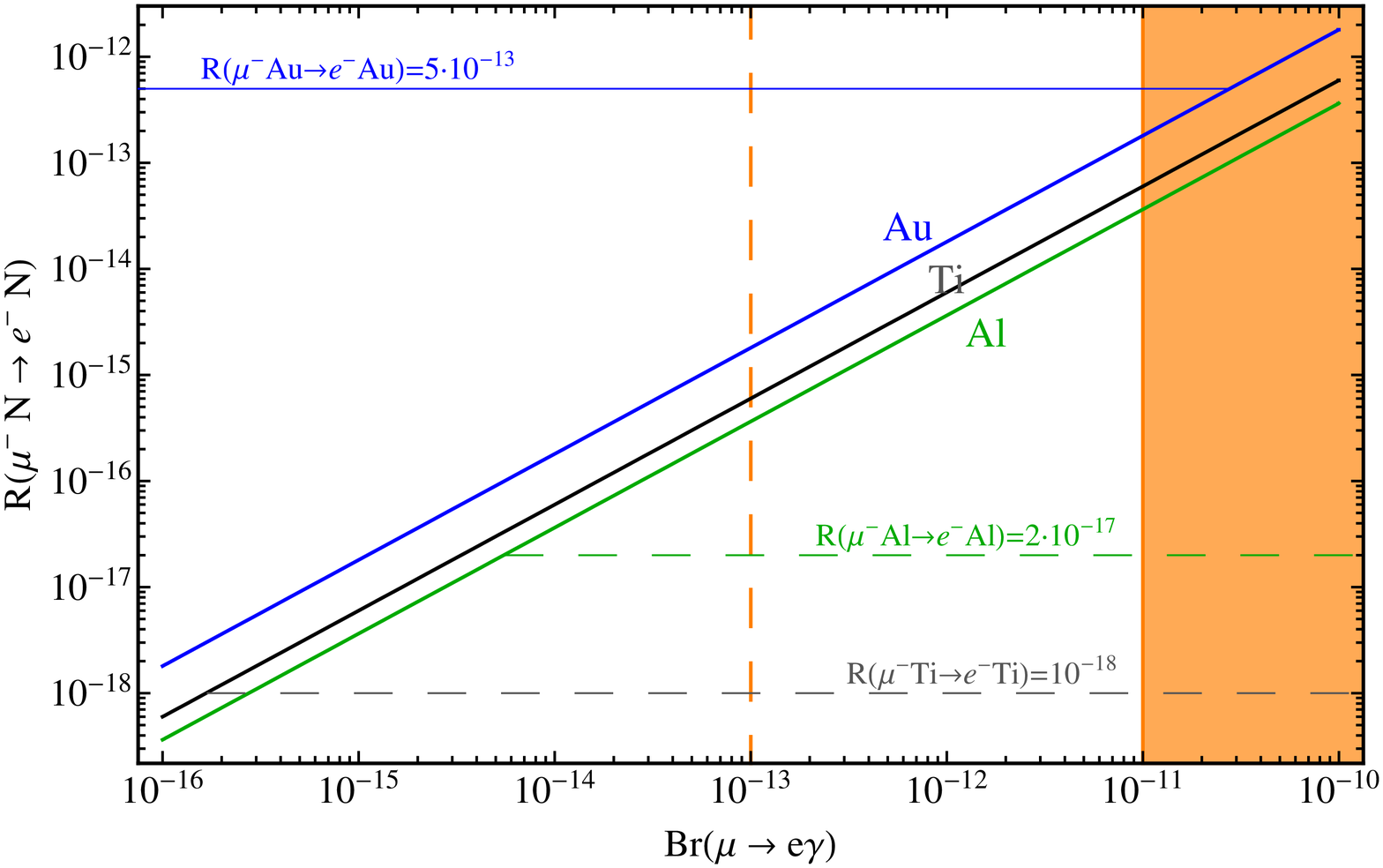}
\caption{Inverse seesaw model expectations for $\mu \to e \gamma$
  versus $\sin^2\theta_{13}$ (left panel), and correlation of $\mu \to
  e \gamma$ with $\mu^--e^-$ conversion rates (right panel), details
  in Ref.~\cite{Deppisch:2004fa}.}
\label{fig:lfv-inverse}
\end{figure} 
The quasi-Dirac neutral heavy leptons present in such extended seesaw
models may mediate large LFV even in the absence of supersymmetry
and,if they have masses around TeV or so, may be directly produced at
accelerators~\cite{Dittmar:1990yg}.

\section{Theory of neutrino mass}
\label{sec:origin-neutrino-mass}

Despite the great experimental progress neutrino physics has recently
undergone, the ultimate origin of neutrino mass remains one of the
most well kept secrets of nature. Here I will not dwell upon the
various options to endow neutrinos with mass, however I will mention
the main broad options.

\subsection{How do neutrinos get mass?}
\label{sec:mechanism}

\begin{figure}[h] \centering
  \includegraphics[ scale=.4]{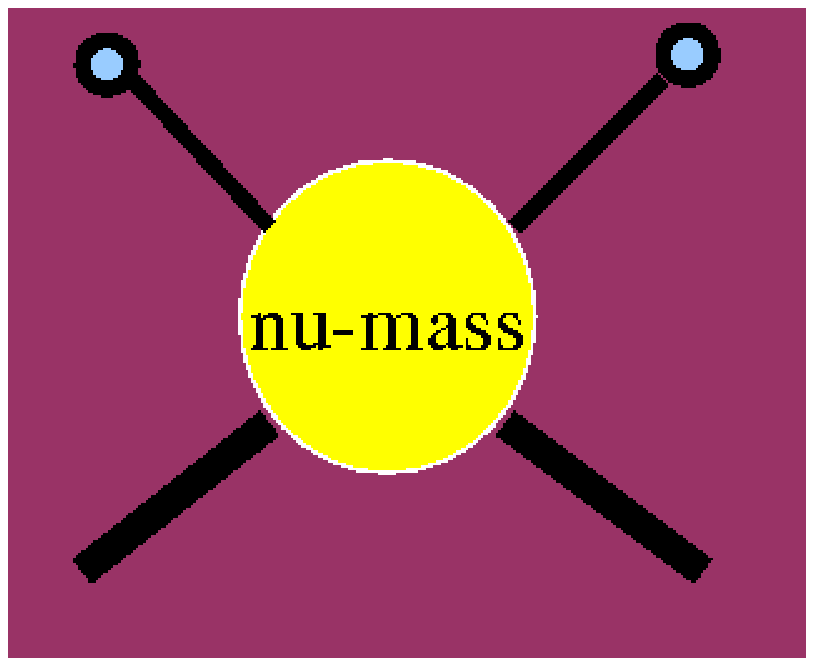} 
\hglue 2cm 
\includegraphics[scale=.62]{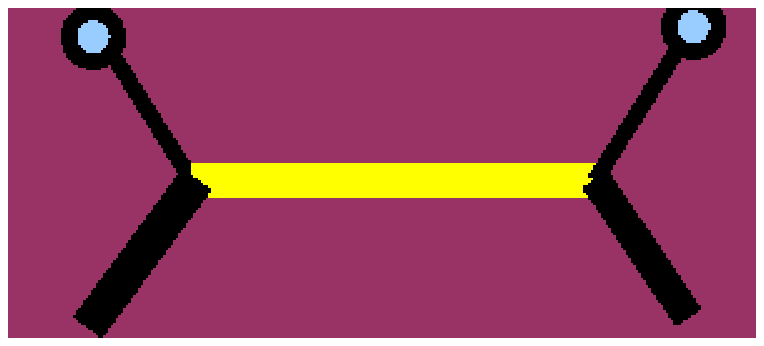} 
\caption{\label{fig:d-5} Dimension-5 operator responsible for
  neutrino mass~\cite{Weinberg:1980bf} and its type-I seesaw
  realization.}
\end{figure}

First note that charged fermions in the Standard Model (SM) come in
two chiral species so that they acquire mass when the electroweak
symmetry breaks through the nonzero vacuum expectation value (vev) of
the Higgs scalar doublet $\vev{\Phi}$.  Neutrinos do not.  There is,
however, an effective lepton number violating dimension-five operator
$\lambda L \Phi L \Phi$ in Fig.~\ref{fig:d-5}, which can be added to
the SM (here $L$ denotes any of the lepton
doublets)~\cite{Weinberg:1980bf}.
This induces Majorana neutrino masses quadratic in the Higgs vev
$\vev{\Phi}$ (indicated by small blobs in Fig.~\ref{fig:d-5}), in
contrast to the linear behavior of the charged fermion masses. This
would provide a natural way to account for the smallness of neutrino
masses, irrespective of their specific origin.
Little more can be said from first principles about the \emph{
  mechanism} giving rise to this operator, its associated mass \emph{
  scale} or its \emph{flavour structure}.  Its strength $\lambda$ may
be suppressed by a large scale $M_X$ in the denominator (top-down)
scenario, leading to $ m_{\nu} = \lambda_0 \frac{\vev{\Phi}^2}{M_X}, $
where $\lambda_0$ is some unknown dimensionless constant.

Since gravity has been argued to break global symmetries, it could
induce the dimension-five operator, with $M_X = M_P$, the Planck
scale~\cite{deGouvea:2000jp}. However in this case the resulting
Majorana neutrino masses are too small, and hence one needs physics
beyond the Standard Model to account for current data.

\begin{figure}[ht] 
     \includegraphics[angle=0,scale=.5]{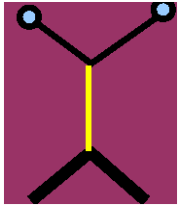}
     \caption{\label{fig:seesaw} %
       Type-II seesaw
       mechanism~\cite{schechter:1980gr}.}
\end{figure}

A popular way to to generate the dimension-5 operator by the exchange
of heavy states, typically fermions (type-I seesaw) as illustrated in
Fig.~\ref{fig:d-5}, right. However, also heavy scalars (type-II
seesaw) can do the job, as shown in Fig.~\ref{fig:seesaw}.
The so-called type-I seesaw mechanism was first mentioned in
Ref.~\cite{Minkowski:1977sc} while the type-II seesaw mechanism was
first mentioned in Ref.~\cite{schechter:1980gr} as part of the general
classification of neutrino mass-giving schemes. The hierarchy of vevs
was discussed in Ref.~\cite{schechter:1982cv}, along with the detailed
perturbative seesaw diagonalization method.
The main point is that, as the masses of the intermediate states go to
infinity, neutrinos become light. The seesaw provides a simple
realization of Weinberg's dimension-5
operator~\cite{Weinberg:1980bf}. It can be implemented in many ways,
with explicitly or spontaneously broken B-L, gauged or not; with
different gauge groups and multiplet contents, minimal or not; with
its basic scale large or small.
All of this, the original references, together with many other
variants of the seesaw
mechanism~\cite{mohapatra:1986bd,gonzalez-garcia:1989rw,Akhmedov:1995vm,Malinsky:2005bi}
are reviewed in \cite{Valle:2006vb,Nunokawa:2007qh}.
Detailed model-independent aspects of seesaw phenomenology, e.~g. the
structure of its lepton mixing matrix are given in~\cite{schechter:1980gr}.

\begin{figure}[h] \centering
    \includegraphics[height=2cm,width=.45\linewidth]{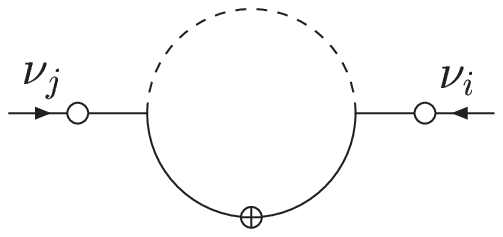}
  \includegraphics[height=3cm,width=.45\linewidth]{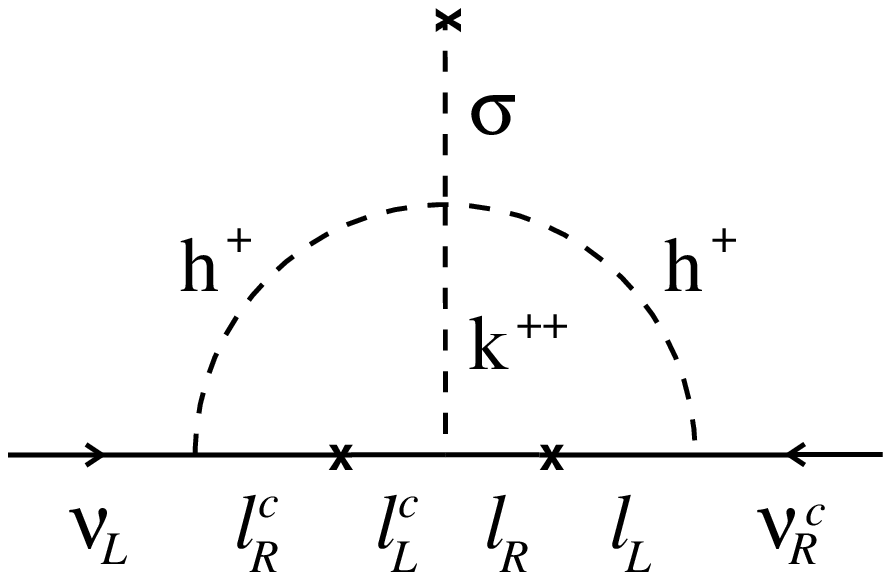}
    \caption{\label{fig:neumass} 
    Radiative origin for neutrino mass. }
\end{figure}

Alternatively, $\lambda$ could vanish due to
symmetry~\cite{Gogoladze:2008wz} or be suppressed by small parameters
(e.g. scales, Yukawa couplings) and/or loop-factors (bottom-up
scenario) with no need for a large scale, opening the door to new
processes associated with the new states required to provide the
neutrino mass and which could be searched for, e.~g., at the LHC.
This is the case in radiative schemes, where neutrino masses arise as
calculable loop corrections~\cite{zee:1980ai,babu:1988ki} as
illustrated in Fig.~\ref{fig:neumass}.
The field $\sigma$ in the right panel is an \321 singlet whose vev
breaks lepton number and induces the neutrino masses, as in
Ref.~\cite{Peltoniemi:1993pd}.  Clearly, in this case neutrino masses
are suppressed by a product of three small Yukawas, two charged lepton
masses, in addition to the two-loop factor.

An example of hybrid neutrino masses is provided by supersymmetry,
which gives a plausible and experimentally testable origin for
neutrino mass.
Indeed, the intrinsically supersymmetric way to break lepton number is
to break the so-called R parity. This may happen spontaneously, driven
by a nonzero vev of an \321 singlet
sneutrino~\cite{Masiero:1990uj,romao:1992vu,romao:1997xf}, leading to
an effective model with bilinear violation of R
parity~\cite{Diaz:1998xc,Hirsch:2004he}. This provides the minimal way
to add neutrino masses to the MSSM~\cite{Hirsch:2004he}. Neutrino mass
generation is hybrid, with typically a normal hierarchy mass spectrum
where one scale (atmospheric) is generated at tree level through a
\emph{weak-scale seesaw} and the other (solar) is induced by
\emph{calculable} one-loop corrections. This is illustrated in the
left panel of Fig.~\ref{fig:neumass}, where open blobs denote the
$\Delta L=1$ insertions present in the R-parity
MSSM~\cite{Hirsch:2000ef}.

Such low-scale models of neutrino mass offer the tantalizing
possibility of reconstructing neutrino mixing at high energy
accelerators, like the LHC and the ILC.  A clear example is provided
by models where supersymmetry is the origin of neutrino mass.
A general feature of these models is that, unprotected by any
symmetry, the lightest supersymmetric particle (LSP) is expected to
decay inside the detector~\cite{Hirsch:2000ef}~\cite{deCampos:2005ri}.
More strikingly, its decay properties correlate with the neutrino
mixing angles. For example, if the LSP is the lightest neutralino, it
should have the same decay rate into muons and taus, since the
observed atmospheric angle is close to
$\pi/4$~\cite{Porod:2000hv,romao:1999up,mukhopadhyaya:1998xj}.
Such correlations hold irrespective of which supersymmetric particle
is the lightest~\cite{Hirsch:2003fe} and constitute a smoking gun
signature of this proposal that will be tested at upcoming
accelerators.

\subsection{How to understand mixing angles?}
\label{sec:understanding-mixing}

Current neutrino oscillation data indicate the existence of two large
lepton mixing angles, while quark mixing angles are all small. This is
rather difficult to \emph{explain} from first principles in unified
schemes where quarks and leptons are related. Phenomenologically,
there seems to be an intriguing complementarity between quark and
lepton mixing
angles~\cite{Raidal:2004iw,minakata-2004-70,Ferrandis:2004vp,Dighe:2006zk}.
There have been many attempts to understand the values of the leptonic
mixing angles from underlying symmetries, a major challenge facing
model-builders.

Harrison, Perkins \& Scott noted~\cite{Harrison:2002kp} that the
neutrino mixing angles are approximately given by,
\begin{align}
\label{eq:hps}
\tan^2\theta_{\Atm}&=\tan^2\theta_{23}^0=1\\ \nonumber
\sin^2\theta_{\textrm{Chooz}}&=\sin^2\theta_{13}^0=0\\
\tan^2\theta_{\Sol}&=\tan^2\theta_{12}^0=0.5 .\nonumber
\end{align}
Such pattern could result from some flavour symmetry valid at high
energy scales.  Its predictions should then be corrected by
renormalization group
evolution~\cite{Altarelli:2005yp,Hirsch:2006je,Altarelli:2004za}.

Here I consider a specific idea to predict neutrino masses and mixing
angles: that neutrino masses arise from a common seed at some
``neutrino mass unification'' scale $M_X$~\cite{chankowski:2000fp},
very similar to the merging of the SM gauge coupling constants at high
energies due to supersymmetry~\cite{amaldi:1991cn}.
Although in its simplest form this idea is now inconsistent (at least
if CP is conserved) with the observed value of the solar mixing angle
$\theta_{12}$, there is an alternative realization in terms of an
$A_4$ flavour symmetry which is both viable and
predictive~\cite{babu:2002dz}. Starting from three-fold degeneracy of
the neutrino masses at the seesaw scale, the model predicts maximal
atmospheric angle and vanishing $\theta_{13}$,
$$\theta_{23}=\pi/4~~~\rm{and}~~~\theta_{13}=0\:.$$ 
Although the solar angle $\theta_{12}$ is unpredicted, one
expects~\footnote{There have been realizations of the $A_4$ symmetry
  that also predict the solar angle, e.~g.
  Ref.~\cite{Hirsch:2005mc}.}
$$\theta_{12}={\cal O}(1).$$ 
If CP is violated $\theta_{13}$ becomes arbitrary and the Dirac phase
is maximal~\cite{Grimus:2003yn}.  One can show that lepton and slepton
mixings are closely related and that there must exist at least one
slepton below 200 GeV, which can be produced at the LHC. The absolute
Majorana neutrino mass scale $m_0 \geq 0.3$ eV ensures that the model
will be probed by future cosmological tests and $\beta\beta_{0\nu}$
searches.  Rates for lepton flavour violating processes $l_j \to \l_i
+ \gamma$ typically lie in the range of sensitivity of coming
experiments, with BR$(\mu \to e \gamma) \gsim 10^{-15}$ and BR$(\tau
\to \mu \gamma) > 10^{-9}$.



\section{Neutrinos as cosmological probes}
\label{sec:neutr-as-cosm}

Optical telescopes can only probe the recent epochs of the evolution
of the Universe, after their last scattering surface, about 400.000 yr
after the primordial Bang. In contrast neutrinos can probe much
earlier stages in the evolution of the early Universe, such as
nucleosynthesis, and even earlier periods.
Indeed, if neutrino masses are generated at high scale, early enough
in the evolution of the Universe, \emph{a la seesaw}, previous to the
electroweak phase transition, they may provide the seed both for the
observed baryon asymmetry of the Universe, and the dark matter, as I
now discuss.

\subsection{Thermal leptogenesis}
\label{sec:thermal-leptogenesis}

\begin{figure}[h]
\centering
\includegraphics[clip,height=2.5cm,width=0.8\linewidth]{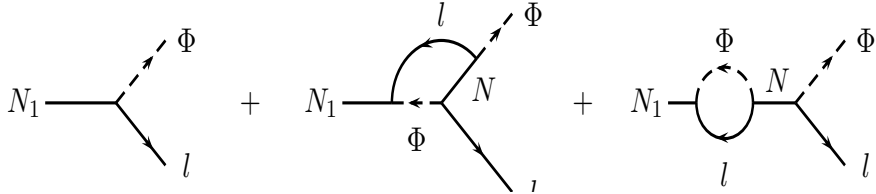}
\caption{Diagrams contributing to  leptogenesis.}
     \label{fig:lep-g}
\end{figure}

\begin{figure}[h]
\centering
\includegraphics[clip,height=4.5cm,width=0.7\linewidth]{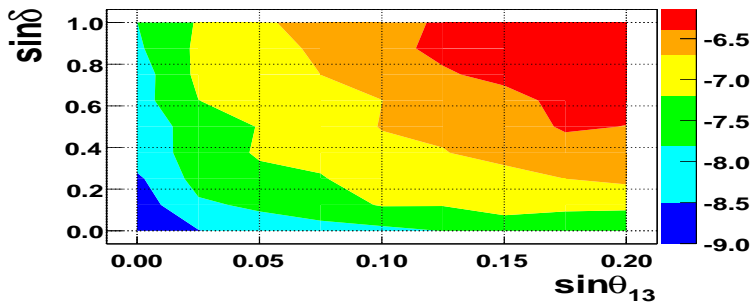}
\caption{Sizeable leptogenesis in supersymmetric SO(10) models, from
  \cite{Romao:2007jr}.}
     \label{fig:lg10}
\end{figure}

Seesaw models open an attractive possibility of accounting for the
observed cosmological matter-antimatter asymmetry in the Universe
through leptogenesis~\cite{Fukugita:1986hr}.
In this picture the decays of the heavy ``right-handed'' neutrinos
present in the seesaw take place before the electroweak phase
transition~\cite{kuzmin:1985mm} through diagrams in
Fig.~\ref{fig:lep-g}. They may also violate CP with rates smaller than
the Hubble expansion rate at that epoch.
Under these circumstances, the lepton (or B-L) asymmetry thus produced
gets converted, through sphaleron processes, into the observed baryon
asymmetry.
However, in typical supersymmetric seesaw schemes the high temperature
needed for leptogenesis leads to an overproduction of gravitinos,
which destroys the standard Big Bang Nucleosynthesis (BBN)
predictions. This happens in minimal supergravity models, with
$m_{3/2} \sim$ 100 GeV to 10 TeV, where gravitinos decay during or
after BBN. 
To prevent such \emph{gravitino crisis} one requires an upper bound on
the reheat temperature $T_R$ after inflation, since the abundance of
gravitinos is proportional to $T_R$. This leads to a stringent upper
bound~\cite{Kawasaki:2004qu}, which is in conflict with the
temperature required for leptogenesis, $T_R > 2 \times 10^9$
GeV~\cite{Buchmuller:2004nz}. One way to cure this
conflict~\cite{Farzan:2005ez} is to add a small R-parity violating
$\lambda_i \hat{\nu^c}_i \hat{H}_u \hat{H}_d$ term in the
superpotential, where $\hat{\nu^c}_i$ are right-handed neutrino
supermultiplets. One can show that in the presence of this term, the
produced lepton-antilepton asymmetry can be enhanced.
An alternative suggestion~\cite{Hirsch:2006ft} was made in the context
of extended SO(10) supersymmetric seesaw schemes. It was shown in this
case that leptogenesis can occur at relatively low scales, TeV or so,
through the decay of a new singlet, thereby avoiding the gravitino
crisis. Washout of the asymmetry is effectively suppressed by the
absence of direct couplings of the singlet to leptons.
The presence of extra chiral singlets also helps to reconcile the
large lepton mixing angles with small quark mixing angles, within the
framework of the successful Fritzsch ansatz.  As illustrated in
Fig.~\ref{fig:lg10} sizeable asymmetry can be generated just from the
leptonic CP violation parameter $\delta$ that characterizes neutrino
oscillations. 

\subsection{Neutrino masses and dark matter}
\label{sec:neutrino-masses-dark}

\begin{figure}[h]
\centering
\includegraphics[clip,height=4.5cm,width=0.45\linewidth]{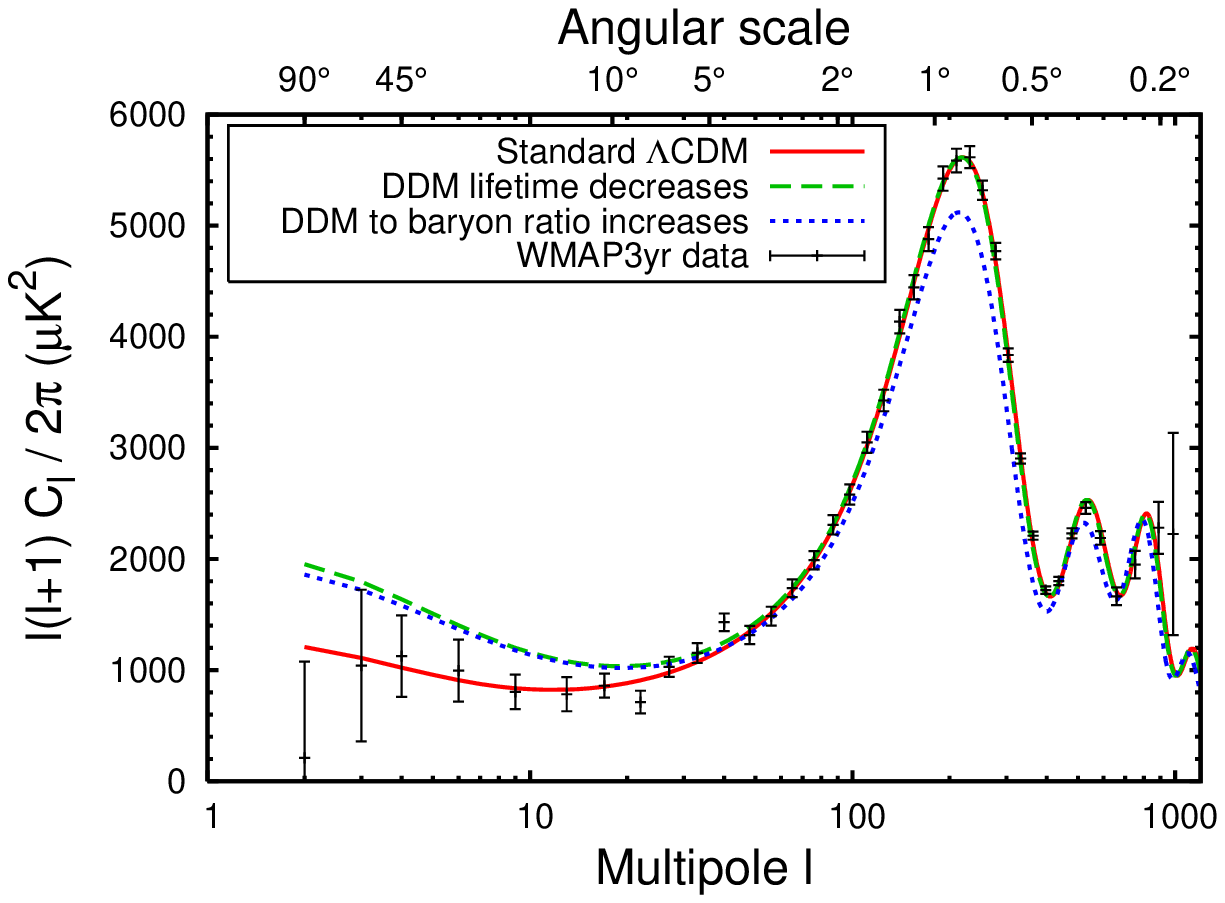}
\includegraphics[clip,height=4.5cm,width=0.45\linewidth]{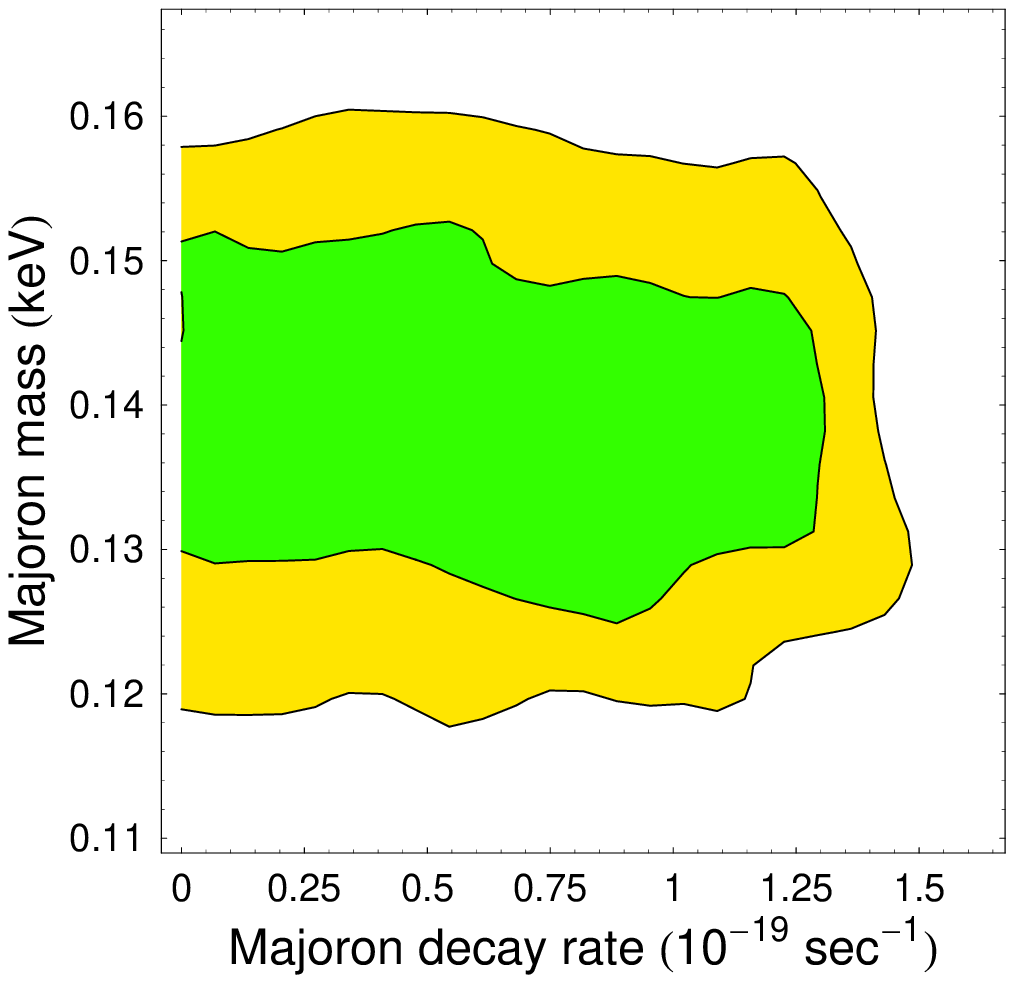}
\caption{Late decaying majoron dark matter parameters, from
  \cite{Lattanzi:2007ux}.}
     \label{fig:kev-maj-cmb}
\end{figure}

\begin{figure}[h]
\centering
\includegraphics[clip,height=4.5cm,width=0.45\linewidth]{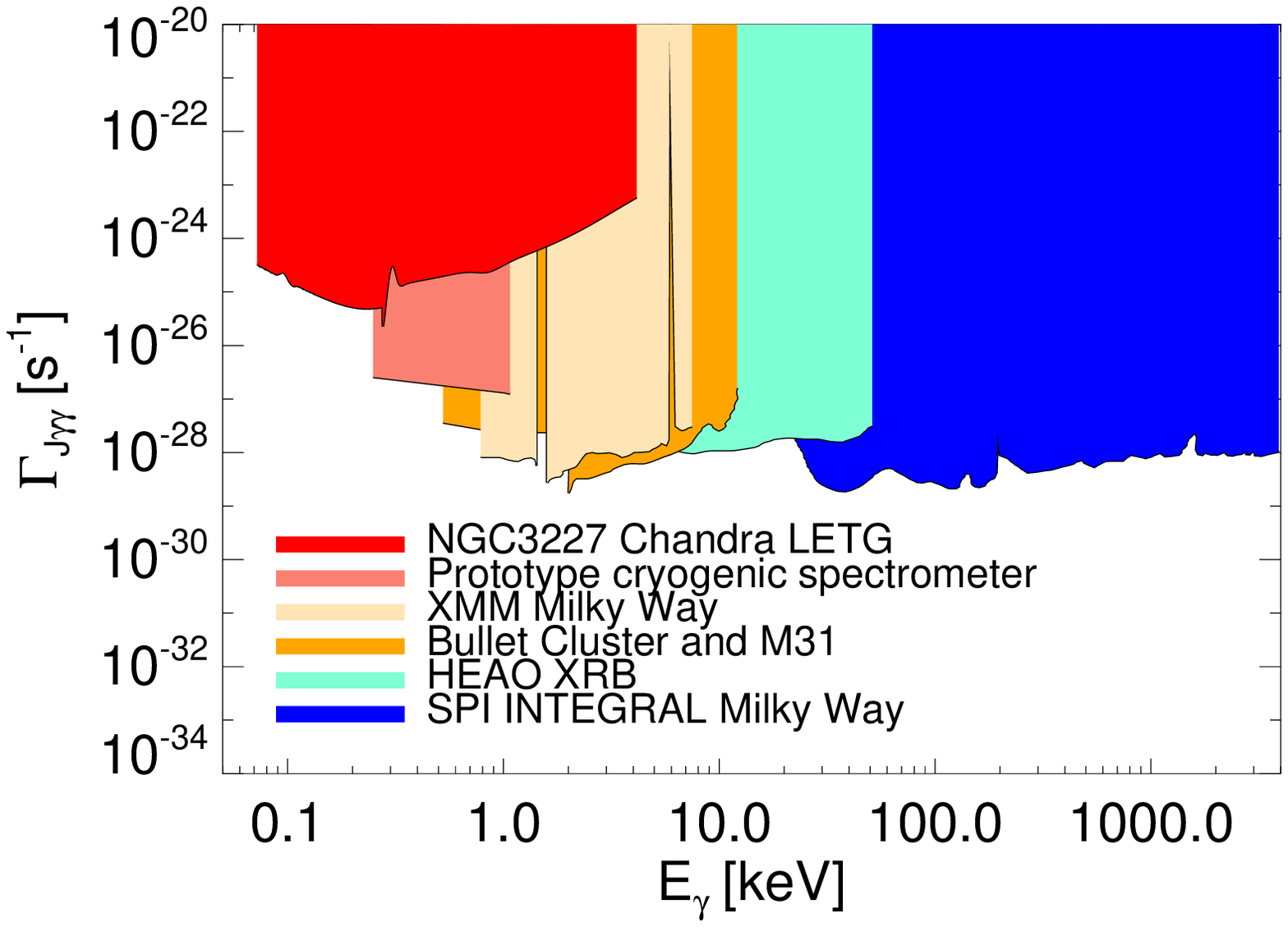}
\includegraphics[clip,height=4.5cm,width=0.45\linewidth]{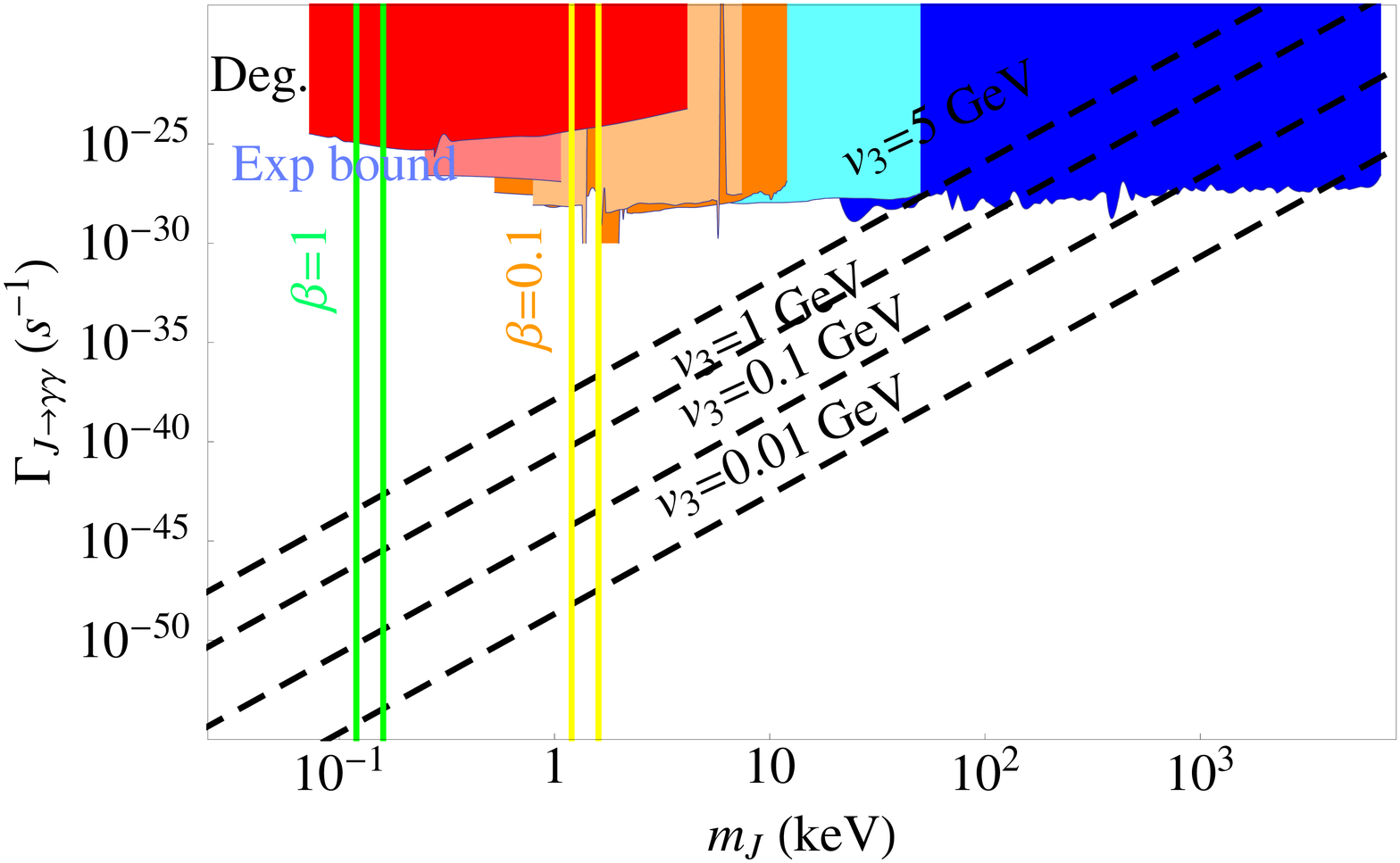}
\caption{Testing the majoron late decaying dark matter, from
  \cite{Bazzocchi:2008fh}.}
     \label{fig:kev-maj-gg}
\end{figure}

An attractive way to generate neutrino masses as required to account
for current neutrino oscillation data involves the spontaneous
breaking of lepton number.
Due to quantum gravity effects the associated Goldstone boson
- the majoron - is likely to pick up a mass.
If its mass lies in the kilovolt scale, the majoron can play the role
of late-decaying Dark Matter, decaying mainly to neutrinos.
However, cosmic microwave background observations place
constraints~\cite{Lattanzi:2007ux}, on the majoron lifetime and mass,
illustrated in Fig.~\ref{fig:kev-maj-cmb},
Such majoron decaying dark matter scenario fits nicely in models where
neutrino masses arise \emph{a la seesaw}, where the majoron couples to
photons due to the presence of a Higgs triplet. In this case it may be
\emph{tested} as it has also a sub-dominant decay to two photons
leading to a mono-energetic emission line.  Comparison of expected
photon emission rates with observations leads to model independent
restrictions on the relevant parameters, illustrated in
Fig.~\ref{fig:kev-maj-gg}
 the resulting sensitivities
within an explicit seesaw realization, 

Finally, I also mention that neutrino mass may open new possibilities
for ``conventional'' supersymmetric dark matter, widely discussed
here. It can be shown~\cite{Arina:2008bb} that, within the inverse
seesaw mechanism for generating neutrino
masses~\cite{mohapatra:1986bd} minimal supergravity is more likely to
have a \emph{sneutrino} as the lightest superparticle than the
conventional neutralino. Such schemes naturally reconcile the small
neutrino masses with the correct relic sneutrino dark matter abundance
and accessible direct detection rates in nuclear recoil
experiments~\cite{Arina:2008bb}.

\vspace{.3cm}

{\em Acknowledgments:} 

I thank the organizers for hospitality and Martin Hirsch for reading
the manuscript. This work was supported by Spanish grants
FPA2008-00319/FPA2005-01269 and FPA2008-01935-E/FPA, as well as
European commission RTN Contract MRTN-CT-2004-503369 and ILIAS/N6
Contract RII3-CT-2004-506222.

\def\baselinestretch{1}

\bibliographystyle{h-physrev4}

\end{document}